\documentclass[prd,twocolumn,showpcs,amsmath,amssymb]{revtex4}

\usepackage{graphicx}
\usepackage{dcolumn}
\usepackage{bm}
\usepackage{color}%
\usepackage{upgreek}
\newcommand{\lb}{\left(}
\newcommand{\rb}{\right)}
\newcommand{\lsb}{\left[}
\newcommand{\rsb}{\right]}

\newcommand{\const}{\mbox{const}}

\newcommand{\bJ}{\bm{J}}
\newcommand{\DD}{\Delta^2}

\usepackage[latin1]{inputenc}

\begin{document}






\title{
Fluid mixing by curved trajectories of microswimmers}

\author{Dmitri O. Pushkin}
\affiliation{The Rudolf Peierls Centre for Theoretical Physics, 1 Keble Road, Oxford, OX1 3NP, UK}
\email[]{mitya.pushkin@physics.ox.ac.uk}

\author{Julia M. Yeomans}
\affiliation{The Rudolf Peierls Centre for Theoretical Physics, 1 Keble Road, Oxford, OX1 3NP, UK}
\email[]{j.yeomans1@physics.ox.ac.uk}
\homepage[]{http://www-thphys.physics.ox.ac.uk/people/JuliaYeomans/}


\date{\today}

\begin{abstract}
We consider the  tracer diffusion $D_{rr}$ that arises from the run-and-tumble motion of low Reynolds number swimmers, such as bacteria.  Assuming a dilute suspension, where the bacteria move in uncorrelated runs of length $\lambda$, we obtain an exact expression for $D_{rr}$ for dipolar swimmers in three dimensions, hence explaining the surprising result that this  is independent of $\lambda$. 
We compare $D_{rr}$ to the contribution to tracer diffusion from entrainment.
\end{abstract}



\maketitle

As microswimmers, such as bacteria, algae or active colloids, move they produce long-range velocity fields which stir the surrounding fluid. As a result particles and biofilaments suspended in the fluid diffuse more quickly, thus helping to ensure an enhanced nutrient supply.  Following the early studies of mixing in concentrated microswimmer suspensions \cite{Wu00,Dombrowski04,Sokolov09}, recent experiments have demonstrated enhanced diffusion in dilute suspensions of {\em Chlamydomonas reinhardtii}, {\em Escherichia coli} and self-propelled particles \cite{Leptos,Kurtuldu,Mino11,Mino12,Poon13}. Simulations 
have found similar behaviour \cite{Underhill08,IshiPedley10,LinChildress11} and microfluidic devices exploiting the enhanced transport due to motile organisms have been suggested \cite{KimBreuer04,Kimbreuer07}.  However, theoretical description of fluctuations and mixing in active systems remains a challenge even for very dilute suspensions of microswimmers.  The statistics of fluid velocity fluctuations was studied in \cite{Rushkin,Zaid11,Underhill11}. As the tracer displacements at short times are proportional to fluid velocities, these results characterise the short-time statistics of tracer displacements. In particular, the fluid velocity fluctuations turn out, generically, non-Gaussian. Features of the long-time tracer displacement statistics remain unknown. 


The Reynolds number associated with bacterial swimming is $\sim 10^{-4}-10^{-6}$. Therefore the flow fields that result from the motion obey the Stokes equations and the far velocity field can be described by a multipole expansion. The leading order term in this expansion, the Stokeslet (or Oseen tensor), which decays with distance $\sim r^{-1}$, is the flow field resulting from a point force acting on the fluid. However biological swimmers, which are usually sufficiently small that gravity can be neglected, move autonomously and therefore have no resultant force or torque acting upon them. Hence the Stokeslet term is zero and the flow field produced by the microswimmers contains only higher order multipoles, for example dipolar contributions, $\sim 1/r^2$, and quadrupolar terms, $\sim 1/r^3$.

\begin{figure}
\includegraphics[width=0.89\columnwidth]{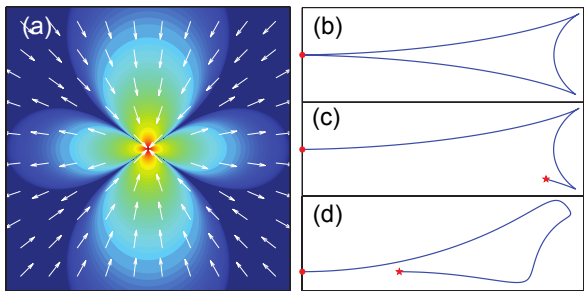}
\caption{ (a) The angular dependence of the dipolar flow field. The velocity decays as $r^{-2}$ where $r$ is the distance from the swimmer. (b) A typical, closed-loop tracer trajectory for an infinite, straight swimmer path and tracer velocity $\ll$ swimmer velocity. (c) A typical trajectory for a finite swimmer path (d) A typical entrained trajectory, for an infinite swimmer path, and tracer close to the swimmer.
}
\label{f:tracer_paths}
\end{figure}

The absence of the Stokeslet term has important repercussions for the  way in which tracer particles are advected by swimmers. The angular dependences of the dipolar velocity field -- shown in Fig.~\ref{f:tracer_paths} -- and of  higher order multipoles of the flow field lead to loop-like tracer trajectories. For a distant swimmer, moving along an infinite straight trajectory these loops are closed.

The paths of bacteria or active colloids are, however, far from infinite straight lines. For example, periodic tumbling (abrupt and substantial changes in direction) is a well established mechanism by which microorganisms such as {\it E-coli} can move preferentially  along chemical gradients. Even in the absence of tumbling, microswimmers typically have curved paths due to rotational diffusion or non-symmetric swimming strokes. For non-infinite swimmer trajectories tracers no longer move in closed loops and the swimmer reorientations cause enhanced diffusion \cite{LinChildress11}. In this letter we obtain an exact expression for the diffusion constant $D_{rr}$ due to uncorrelated random reorientations of dipolar swimmers in three dimensions. The result allows us to explain the surprising observation \cite{LinChildress11} that, for dipolar swimmers in 3D, the diffusivity is independent of the swimmer run length. 
We then extend our results to give scaling arguments for $D_{rr}$ for swimmers confined to general dimensions $d$ (but retaining the 3D nature of the flow field), and a flow field that decays as $r^{-m}$ where, for example, $m=2$ corresponds to dipolar swimmers such as {\em E. Coli} and {\em C. reinhardtii}; $m=3$ to quadrupolar swimmers such as {\em Parmecium} and to active colloids. 
We find two distinct regimes: for $d>d^{*}(m)=2(m-1)$  the distribution of tracer path lengths converges to a Gaussian \cite{Pushkin13c}; however, for $d<2(m-1)$  the tracer path length distribution takes a truncated Levy form. We compare $D_{rr}$ to another contribution to tracer diffusion, entrainment by the swimmers.

We first calculate $D_{rr}$ for dipolar swimmers ($m=2$) for $d=3$. We consider a swimmer moving along a straight line segment of a trajectory, of length $\lambda$, from an initial position $i$ to a final position $f$. For dipolar swimmers with velocity $\bm{V}=V \bm{k}, \, \| \bm{k} \|=1$ the leading order term in the far-field expansion of the velocity is the stresslet
\begin{equation}
\bm{v} (\bm{r})\approx - \kappa \, \bm{k} \cdot (\bm{k} \cdot \nabla)  J(\bm{r}),
\label{eq:stresslet}
\end{equation}
 where $\bm{r}$ is the radius-vector with the origin at the swimmer, $\kappa$ is the swimmer dipole strength, and
$J(\bm{r})$ is the Oseen tensor,
\begin{equation}
J(\bm{r})=\frac{\bf{I}}{r}+\frac{\bm{r}\bm{r}}{r^3}.
\end{equation}
(In this notation the fluid viscosity and numerical constants are adsorbed in $\kappa$.)  The swimmer velocity fields are extensile for $\kappa >0$ (e.g. for `pushers' such as {\em E. Coli}) and contractile for $\kappa <0$ (e.g. for `pullers' such as {\em C. reinhardtii}).
The validity of the expression (\ref{eq:stresslet}) is based on the assumption that the resultants of the drag and the propulsive forces are parallel to the swimming direction \cite{Pushkin13a} as has been verified experimentally for several types of biological microswimmers \cite{Drescher10,Drescher11}.

We define elliptic co-ordinates, with the origin at the mid-point of the swimmer trajectory and the left and right foci at the initial and final positions of the swimmer, respectively:
\begin{eqnarray}
&x& = a_1 \, \cosh \mu \, \cos  \nu = a_1 \, \sigma  \, \tau, \\
&y& = a_1 \, \sinh \mu \, \sin  \nu, \;
y^2 = a_1^2 \, (\sigma^2-1) \, (1-\tau^2) ,
\end{eqnarray}
where $a_1=\lambda/2$, $\mu \ge 0$, $0 \le \nu \le 2 \, \pi$, $-1 \le \tau \le 1$, and $\sigma \ge 1$. The advantage of using elliptic coordinates is that the square of the tracer displacement can be expressed as a rational function of $\sigma$ and $\tau$ which allows analytical treatment of contributions of different relative positions of the tracer and the swimmer path segments:
\begin{eqnarray}
\Delta^2 &=& \lsb \frac{\kappa}{V} \bm{k} \cdot J \bigg|^f_i \rsb^2=
\lb \frac{\kappa}{V} 
\rb^2 \frac{4}{a_1^2 \, (\sigma^2-\tau^2)^4 } \, \nonumber\\
&&\lsb
   \lb 1-3 \, \tau^2 \rb ^2 \sigma^4 
   + \lb 3 \, \tau^2 -1 \rb   \lb \tau^2 +2 \tau -1 \rb   \right. \nonumber \\ 
 && \left. \lb \tau^2 -2 \tau -1  \rb   \sigma^2 
    + \tau^2 \lb  1 + \tau^2 \rb^2
\rsb .
\label{e:D3_elliptic}
\end{eqnarray}

To obtain the diffusion coefficient we assume a uniform isotropic distribution of swimmers that move in straight runs of length $\lambda$. We assume no interactions between the swimmers and statistical independence of the consecutive runs. Then the mean squared displacement of the tracer $< \DD >$ can be obtained by summing over all possible straight segments of swimmer paths
\begin{equation}
< \DD > \, = n_s \, \int d^d \bm{r_i} \, \DD(\bm{r_i})
\label{e:D}
\end{equation}
where $n_s$ is the concentration of the segments and $\bm{r_i}$ the initial distance between the tracer and the swimmer.
When the time $t$ is greater that the reorientation time, each swimmer contributes on average $Vt/\lambda$ segments. Therefore the diffusion coefficient $D_{rr}$ due to uncorrelated, random swimmer reorientations is
\begin{equation}
D_{rr}=< \DD > /\,( 2 \, d \, t) \, =  \frac{1}{2d} \frac{n \, V}{\lambda} \, \int d^d \bm{r_i} \, \DD(\bm{r_i})
\label{e:DD}
\end{equation}
where $n$ is the number density of swimmers. 

Substituting in the expression (\ref{e:D3_elliptic}) for the displacement due to a single swimmer, and transforming to elliptical co-ordinates, 
\begin{equation}
 D_{rr}  = \, n \lambda^d \cdot \frac{V}{\lambda}  \cdot \lb \frac{\kappa}{V \lambda} \rb^2 \cdot I_{d,2} \lb \frac{a}{\lambda} \rb
\label{e:dipScal}
\end{equation}
where the subscripts $d,2$ indicate that we are considering dipolar swimmers in dimension $d$ and $a$ is a short-distance cut-off introduced to regularise the integral.

We have written $D_{rr}$ in the scaling form (\ref{e:dipScal}) to allow generalisation to $m \neq 2$. However we first note that for dipolar swimmers in $d=3$ the integral can be performed explicitly to give
\begin{equation}
I_{3,2} \lb \frac{a}{\lambda} \rb \to  \frac{4 \pi}{3} \quad \mbox{as} \quad a/\lambda \to 0.
\label{e:limit}
\end{equation}
Hence the powers of $\lambda$ in Eq.~(\ref{e:dipScal}) cancel out! We conclude that for three-dimensional suspensions of dipolar swimmers, tumbling of swimmers (or, equivalently, as we discuss later, curvature of swimmers' trajectories)
leads to an effective diffusion coefficient independent of the tumbling length (mean curvature radius) in the limit ($\ref{e:limit}$).
 The value of the prefactor $4 \pi/3 \approx 4.2$ is in good agreement with the result $3.7$ obtained in   simulations \cite{LinChildress11} (particularly given the difficulties of numerical integration as large samples are needed to resolve the heavy tails of the tracer displacement distribution \cite{Pushkin13c}).

We now identify each of the terms on the rhs of Eq.~(\ref{e:dipScal}) and generalise to arbitrary $d$, $m$.
The first term corresponds to the number of swimmers within one flight of length $\lambda$ from the tracer. The second term accounts for the number of statistically independent path segments. The third term corresponds to a characteristic tracer displacement in an interaction with a single path segment of a swimmer (`collision') at distances $O(\lambda)$. The last term, i.e. the function $I_{d,2} \lb a/\lambda \rb$, arises from the re-scaled lower integration limit in (\ref{e:DD}). (Convergence of the integral at the upper limit does not pose problems.) In the critical phenomena language it may be thought of as the scaling function accounting for the influence of the microscale $a$ at the mesoscale $\lambda$. 
For an arbitrary interaction exponent $m$ and dimension $d$ the expression~(\ref{e:dipScal})  is generalised to
\begin{equation}
 D_{rr}  = \, n \lambda^d \cdot \frac{V}{\lambda}  \cdot \lb \frac{\beta_m a^m}{\lambda^{m-1}} \rb^2 \cdot I_{d,m} \lb \frac{a}{\lambda} \rb,
\label{e:Scal}
\end{equation}
where $\beta_m$ is the dimensionless strength of the dominant $m$-pole term of the far velocity field.

\begin{figure}[t]
\label{f:turn}
\includegraphics[width=0.6\columnwidth]{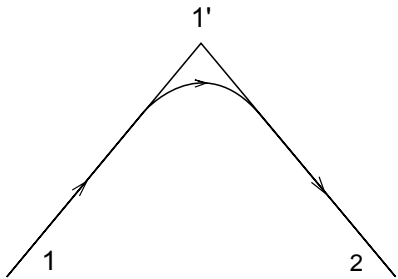}
\caption{A swimmer path with an abrupt and smooth turns.}
\label{f:turn2}
\end{figure}

The asymptotic behaviour of $I_{d,m} \lb x \rb$ as $x \to 0$ depends on the singularities of $\Delta^2$ at the endpoints of the swimmer paths, which correspond to the maximal tracer displacements. Two cases can be distinguished depending on the fundamentally different statistical behaviour of large tracers displacements which scale as $\Delta \sim r^{-(m-1)}$ where $r$ is the distance to the endpoint. For a uniform distribution of swimmers the probability of an endpoint at a distance $r$ from a tracer is proportional to the spherical volume of radius $r$.  Hence the probability distribution of tracer displacements for large $\Delta$
\begin{eqnarray}
p(\Delta) d \Delta \propto d(r^d) \propto d ( \Delta^{-d/(m-1)} ) \propto \Delta^{-1-\alpha} d \Delta,
\end{eqnarray}
where $\alpha= {d}/{(m-1)}.$
The central limit theorem \cite{Pushkin13c} guarantees that the ensuing random walk is Gaussian when the dispersion $\langle \Delta^2\rangle$ 
is finite. This requirement results in condition $d>d_*(m)=2(m-1)$.  (Note that this condition differs from that required for the velocity fluctuations to be Gaussian \cite{Rushkin,Zaid11}, which admitted Stokeslet but excluded the stresslet and higher order multipoles.) In this case, that we will refer to as {\em regular}, $I_{d,m} \lb a/\lambda \rb \to \const$.  (It can also be referred to as asymptotics of the first kind \cite{Barenblatt} or the zero anomalous dimension \cite{Goldenfeld}.) Three-dimensional suspensions of dipolar swimmers give an example of the regular case for which fluctuations caused by all points on the swimmer paths contribute to the tracer diffusion. The resulting diffusion coefficient scales as
\begin{eqnarray}
D_{rr} \sim \beta_m^2 \, n \, V \, \lambda^{d-2m+1} \, a^{2m}, \quad d>d_*(m).
\label{e:Dreg}
\end{eqnarray}

A different behaviour emerges when $d<d_*(m)$: $I_{d,m} \lb a/\lambda \rb \sim (a/\lambda)^{d-2(m-1)}$. In this {\em singular} case the diffusion is dominated by interactions with the turning points of swimmer trajectories and the ensuing random walk can be formally considered as a truncated Levy flight. (This case may also be referred to as asymptotics of the second kind \cite{Barenblatt} or a nonzero anomalous dimension \cite{Goldenfeld}.) The resulting diffusion coefficient
\begin{eqnarray}
D_{rr} \sim \beta_m^2 \, n \, V \, \lambda^{-1} \, a^{d+2}, \quad d<d_*(m).
\label{e:Dsing}
\end{eqnarray}
In the borderline singular case $d=d_*(m)$, $I_{d,m} \lb a/\lambda \rb \sim \log (\lambda/a)$ and 
\begin{eqnarray}
D_{rr} \sim \beta_m^2 \, n \, V \, \lambda^{-1} \, a^{d+2} \log (\lambda/a), \quad d=d_*(m).
\label{e:DsingLog}
\end{eqnarray}
\\
\noindent As $d_*(2)=2$ and $d_*(3)=4$, it follows immediately that quasi-two-dimensional suspensions of dipole swimmers (i.e. suspensions where both the swimmers and the tracers are confined to a surface, while retaining the 3D nature of hydrodynamics interactions) and suspensions of quadrupolar swimmers are singular cases. 

We now pause to show that the effect of sharp reorientations of swimmers on the tracer displacement is equivalent to the effect of smooth, curved swimmer trajectories having a finite persistence length. To this end we show that both effects can be treated on the same theoretical footing. Indeed, the tracer displacement due to the far-field of a swimmer moving on a smooth path can be obtained by 
integrating (\ref{eq:stresslet}) by parts:
\begin{equation}
\Delta \bm{r_t} = -\frac{\kappa}{V} \, \bm{k} \cdot \bm{J} \bigg|^{f}_{i}
+ \frac{\kappa}{V} \, \int_i^f d\bm{k} \cdot \bJ.
\label{e:drt}
\end{equation}
The consistency of this expression can be checked by comparing the resulting $\Delta \bm{r_t}$ for a sharp and a smooth turn depicted on Fig.~\ref{f:turn2}. The tracer displacements calculated according to (\ref{e:drt}) should become identical as the smooth turn approaches the sharp one. For the polygonal curve $1\,1'\,2$, assuming the velocity perturbations due the turning at $1'$ negligible,
\begin{widetext}
\begin{eqnarray}
\bm{\Delta r_t}=  -\frac{\kappa}{V} \lsb 
 \bm{k_1} \cdot \bm{J} \bigg|^{1'}_{1} + \bm{k_2} \cdot \bm{J} \bigg|^{1'}_{2}
 \rsb
= \frac{\kappa}{V} \lsb 
- \bm{k} \cdot \bm{J} \bigg|_{1}^{2} + \lb  \bm{k_2}-\bm{k_1} \rb \bm{J}(1')
\rsb.
\end{eqnarray}
\end{widetext}
This result clearly concurs with (\ref{e:drt}) for the curved trajectory $1\rightarrow2$. Thus, the effects of swimmer path curvature can be fully understood by approximating swimmer trajectories by straight runs with sharp re-orientations.


Reorientation of swimmer trajectories is just one factor leading to enhanced mixing in swimmer suspensions. Another contribution is mixing by entrainment \cite{Pushkin13a}
where the tracer is pulled along by the swimmer motion as shown in Fig.~\ref{f:tracer_paths}. This mechanism is due to the Lagrangian contribution to the tracer velocity field, and is a mathematically distinct mechanism for non-closure of tracer loops. 
As swimmer reorientations and tracer entrainment are mathematically and  physically distinct, it is reasonable to assume as a first approximation that their effects add up and the total diffusion
\begin{eqnarray}
D \approx D_{rr} + D_{entr} +D_{therm},
\label{e:Dsum}
\end{eqnarray}
where $ D_{therm}$ is the contribution due to thermal noise, which depends on the physical properties of the diffusing agents, which may be colloidal particles \cite{Leptos}, large polymer molecules \cite{KimBreuer04}, or non-motile bacteria \cite{Poon13}.
While other factors, such as enhanced mixing due to multiple swimmers and their interactions are clearly important for concentrated suspensions of swimmers, they can be neglected for dilute suspensions. Also note that correlations of swimmer path segments, neglected by the current model, may influence the diffusion.

We now compare rates of mixing due to random reorientations and due to entrainment.
$D_{entr}$ can be estimated using a generalised kinetic theory argument \cite{LinChildress11,Pushkin13a}:
\begin{eqnarray}
D_{entr} \approx \frac{1}{2d} V \, l \, n \, v_{entr},
\end{eqnarray}
where $l$ is the entrainment length, and $v_{entr}$ is the volume of fluid entrained by an individual swimmer. Assuming that both the short-distance cut-off and the entrainment length are of the order of the swimmer physical size and that the entrained volume is of the order of the swimmer's own volume, $l \sim a$, $v_{entr} \sim a^d$, and we obtain
\begin{eqnarray}
D_{entr} \sim n \, V \, a^{d+1}.
\label{e:DEntrScaling}
\end{eqnarray}
Comparing this expression with (\ref{e:Dsing}) shows that {\em in the  singular} cases $D_{rr} \ll D_{entr}$ when $\lambda \gg a$.  As the dimensionless swimmer strength $\beta_m=O(1)$, mixing due to entrainment already dominates mixing due to trajectory curvature for $\lambda \gtrsim a$. We conclude that for quasi-two-dimensional suspensions of dipole swimmers and for suspensions of quadrupole swimmers, the diffusion is dominated by mixing due to entrainment. 

For three-dimensional dipolar swimmers
comparing the expressions (\ref{e:Dreg}) and (\ref{e:DEntrScaling}) gives
$D_{rr, dip}^{3D}/D_{entr}^{3D}\approx 6 \beta_2^2$ and the effect of trajectory reorientation overtakes entrainment at $\beta_2 \gtrsim 0.4$. These results are in good agreement with numerical simulations \cite{Pushkin13b,LinChildress11}. 

The experiments \cite{Leptos} examined mixing induced by {\em C. reinhardtii} at time scales much smaller than their reorientation time and found $D_{exper}/\Phi \approx 81 \upmu \!m^2/s$, where $\Phi$ is the volume fraction of swimmers. At these time scales, $D \approx D_{entr}$ and assuming $V=100 \upmu \!m/s$ and $a=5 \upmu \!m$ we find $D/\Phi \approx 83 \upmu \!m^2/s$  in good agreement with the experiments. Assuming $\beta \approx 1$, we predict a $7$-fold increase of the effective diffusion coefficient at longer times. 

In very recent experiments \cite{Poon13}, the tracer diffusion was studied for time scales spanning more than two orders of magnitude in fully three-dimensional suspensions of swimmers, {\em E. Coli}. It was found that $D_{exper}/(n V) = 7 \pm 0.4 \upmu \!m^4$. Using $\beta =1.45 \upmu \!m^2$ obtained by fitting of the flow field of an individual {\em E. Coli} bacterium \cite{Drescher11}, $v_{entr} \approx 1.4 \upmu \!m^3$ and $a \approx 1.4 \upmu \!m$, we obtain the estimate for the total diffusion coefficient $D/(n V) \approx 9 \upmu \!m^4$,
in good agreement with the experiments \cite{Poon13} (and similar estimates therein).



The long-range velocity field of microswimmers gives rise to strong non-equilibrium fluctuations in the surrounding fluid and enhances mixing.
Despite the conceptual appeal of the notion of the `effective temperature' of the bacterial bath \cite{Wu00}, several theoretical approaches to this problem based on extending the fluctuation--dissipation theorem have been shown to fail \cite{Chen2007fluctuations,Underhill11}. The culprits for this failure are long-range spatial and temporal correlations of fluid velocity fluctuations that develop even for very dilute suspensions of uncorrelated swimmers \cite{Underhill11}. While PDFs of short-time tracer displacements are identical to PDFs of fluid velocities and have been considered in the past \cite{Rushkin,Zaid11}, in this work we considered long-time tracer displacements in suspensions of uncorrelated swimmers moving with a finite persistence length $\lambda$. Building up on the model suggested in \cite{LinChildress11}, we first derived an exact expression for the effective diffusion coefficient due to random reorientations in suspensions of 3D dipolar swimmers. We showed that 3D suspensions of dipole swimmers are special in that $D_{rr, dip}^{3D}$ turns out independent of $\lambda$ due to a fortuitous balance between the interaction exponent $m=2$ and the dimension $d=3$. For swimmers with the dimensionless dipolar strength $\beta \approx 1$, $D_{rr, dip}^{3D}$ contributes roughly $85 \%$ of the total induced tracer diffusion, the remaining $15 \%$ are contributed by the entrainment mechanism \cite{Pushkin13a}. We showed that as the Central Limit Theorem holds, the PDF of tracer displacements approaches a Gaussian \cite{Pushkin13c} and that all points on the swimmers trajectories contribute to the effective diffusion coefficient.

The situation is different for quasi-two-dimensional suspensions of swimmers and suspensions of quadrupole swimmers. For them, the induced tracer diffusion is dominated by the entrainment mechanism.

Our results rely on the separation of length scales $a \ll \lambda \ll L$, where $L$ is the macroscopic size of the system. As the mean curvature of the swimmers' trajectories decreases, $\lambda$ eventually becomes comparable to $L$ and the finite system size effects will reduce the effective value of $D_{rr}$. For $\lambda / L \to \infty$, $D_{rr}$ will reduce to zero, while $D_{entr}$ will not change. Thus, the effective diffusion for $\lambda \gg L$ will be dominated by entrainment, in agreement with physical expectations. The other limit, $\lambda \sim a$ may also turn out physically relevant, in particularly when the curvature of swimmers' trajectories is due to strong rotational diffusion. We plan to analyse these effects in the future.


\end{document}